\documentclass[prl,twocolumn,aps,10pt]{revtex4}
\usepackage{graphicx}
\usepackage{amsfonts}
\usepackage{psfrag}
\usepackage{subfigure}
\usepackage{tabularx}
\usepackage{nicefrac}

\usepackage[T1]{fontenc}
\newcommand\cc[1]{#1^{^{\kern-6pt \circ}}\kern2pt}

\def\pa{\partial}
\renewcommand{\a}{\alpha}
\renewcommand{\b}{\beta}

\def\be{\begin{equation}}
\def\ee{\end{equation}}
\def\bea{\begin{eqnarray}}
\def\eea{\end{eqnarray}}
\def\ba{\begin{array}}
\def\ea{\end{array}}
\def\bi{\begin{itemize}}
\def\ei{\end{itemize}}

\newcommand{\beq}{\begin{equation}}
\newcommand{\eeq}{\end{equation}}
\newcommand{\beqn}{\begin{eqnarray}}
\newcommand{\eeqn}{\end{eqnarray}}
\newcommand{\bga}{\begin{align}}

\def\dalemb#1#2{{\vbox{\hrule height .#2pt
\hbox{\vrule width.#2pt height#1pt \kern#1pt
\vrule width.#2pt}
\hrule height.#2pt}}}

\begin{document}

\begin{flushright}
\widetext{CCTP-2011-25, CPHT-RR006.0211}
\end{flushright}

\title{Holographic Three-Dimensional  Fluids with Nontrivial Vorticity}
\parskip = .1cm

\author{ Robert G. Leigh}
\thanks{Perimeter Institute, 31 Caroline St. N., Waterloo ON, Canada and 
Department of Physics, University of Illinois, 1110 W. Green Street, Urbana IL 61801, U.S.A.}
\author{ Anastasios C. Petkou}
\thanks{Department of Physics, University of Crete, GR-71003, Heraklion, Greece}
\author{P. Marios Petropoulos}
\thanks{Centre de Physique Th\'eorique, \'Ecole Polytechnique, CNRS UMR 7644, 91128 Palaiseau Cedex, France}
\preprint{CCTP-2011-25, CPHT-RR006.0211} 

\date{\today}
\begin{abstract}
Three-dimensional fluids with nontrivial vorticity can be described holographically. It is well-known that the Kerr-AdS geometry gives rise to a cyclonic flow. Here we note that Taub--NUT--AdS$_4$ geometries give rise to a rotating fluid with vortex flow.
The  Randers and Zermelo forms of the boundary metrics provide alternative descriptions of the fluid by inertial co-moving or by accelerated observers. Such fluids possess acoustic horizons. Moreover, light propagation on the boundary Taub--NUT fluid will encounter an optical horizon associated with closed timelike curves. In the latter case the Misner string introduces a multi-valuedness of the scalar fluctuations which can be attributed to the anyonic nature of the boundary vortex.

\end{abstract}

\maketitle

The versatile AdS/CFT framework has been recently used for the description of a wide variety of strongly coupled condensed matter systems such as high-$T_\mathrm{c}$ superconductors or superfluids \cite{Hartnoll:2008vx}, strange metals \cite{Faulkner:2010da} and quantum Hall fluids \cite{Dolan:2010mx}. Holographic techniques have been also used for the description of nearly perfect hydrodynamics \footnote{Namely fluids whose viscosity to entropy density ratio is close to the proposed universal ``bound'' $\nicefrac{\eta}{s}\approx\nicefrac{\hbar}{4\pi k_\mathrm{B}}$ \cite{Policastro:2001yc}. It can be shown that all holographic fluids are nearly perfect.}  via a derivative expansion of asymptotically AdS solutions of Einstein's equations \cite{Minwalla1}.  Two interesting classes of  strongly coupled condensed matter systems in 2+1 dimensions which have not been widely discussed in the context of holography include fast-rotating Bose gases and analogue gravity models. These systems are neutral, nevertheless rotation produces effects similar to those of a gauge field. For example, fast rotating atomic gases behave similarly to charged particles in a magnetic field and hence are believed to form a bosonic strongly coupled quantum Hall state for small filling fraction ({\it i.e.}, when the number of vortices is comparable to the number of particles) \cite{cooper:2009}. Also, in analogue gravity systems rotation drags the propagating wave and arguably gives rise to acoustic/optical horizons \cite{Unruh} and Berry phases \cite{Leonhardt}. 

Since the  description of the systems above would involve neutral holographic fluids in nontrivial flows, one needs to understand better the possible equilibrium states in the absence of dissipation before dwelling into the calculations of the physically relevant transport properties. 
In this letter we outline the main results of our study of geometries that are holographically dual to rotating fluids with vorticity. We focus on the Kerr--AdS$_4$ and Taub--NUT--AdS$_4$ geometries. A detailed account will be presented elsewhere \cite{LPP}.

To approach this problem, we find it illuminating to use the $3+1$-split formalism \cite{Leigh:2007wf, Mansi:2008br, Mansi:2008bs}. 
Bulk solutions are taken in the Fefferman--Graham form  
\footnote{The starting point is the Einstein--Hilbert action in the Palatini first-order formulation, with cosmological 
constant $\Lambda=-\nicefrac{3}{ L^2}$ 
and orthonormal co-frame $E^A$, $A=r,a$. The signature is $+-++$, the first direction  $r$ is the holographic one and we will use $a,b,c,\ldots =0,1,2$ for transverse Lorentz indices along with $\alpha,\beta,\gamma=1,2$. Coordinate indices will be denoted $\mu,\nu,\ldots$ and $i,j,k, \ldots$ for transverse spacetime and spatial coordinates respectively, with $\mathrm{x}\equiv(t,x^1,x^2)\equiv(t,x)$.}
\beq
\label{FGform}
\mathrm{d}s^2 = \frac{L^2}{r^2}\mathrm{d}r^2 +\frac{r^2}{L^2}\eta_{ab}E^a(r,\mathrm{x})E^b(r,\mathrm{x})\,.
\eeq
For torsionless connections there is always a suitable gauge choice 
such that the metrics (\ref{FGform}) are fully determined by two coefficients $\hat{\mathrm{e}}^a$ and $\hat{f}^a$ in the 
expansion of the co-frame one-forms $E^a(r,\mathrm{x})$ along the holographic coordinate \bea
\hspace{-.4cm}
&E^a(r,\mathrm{x})= \left[\hat{\mathrm{e}}^a(\mathrm{x})+\frac{ L^2}{r^2} \hat{F}^a(\mathrm{x})+\cdots \right]+\frac{ L^3}{r^3}\left[\hat{f}^a(\mathrm{x})+\cdots \right]&\label{vielbein}
\eea
as it approaches the boundary at   $r\to\infty$.
Other coefficients 
are determined by $\hat{\mathrm{e}}^a$ and $\hat{f}^a$ and have interesting geometrical interpretation \footnote{For example, the coefficient $\hat{F}^a$ is related to the Schouten tensor.}.
The 3+1-split formalism makes clear that $\hat{\mathrm{e}}^a(x)$ and $\hat{f}^a(x)$, viewed now as vector-valued one-forms in the boundary, are the proper  canonical variables playing the role of boundary position and momentum for the radial Hamiltonian evolution. 

We consider here the Kerr--AdS$_4$ and the Lorentzian TN--AdS$_4$ geometries. The first was obtained in \cite{Carter} and we use here the form given e.g. in Eq. (2.1) of \cite{Pope}. The latter geometry was given in Eq. (2.1) of \cite{Mann}. Both bulk geometries give rise to stationary boundary metrics conformal to
the Randers form \cite{Gibbons}
\beq
\label{Randersmetric}
\hat g=-\left( \mathrm{d}t-b_i(x) \mathrm{d}x^i\right)^2+a_{ij}(x) \mathrm{d}x^{i} \mathrm{d}x^{j}\,.
\eeq
The boundary co-frame is taken to have the generic form 
$\hat{\mathrm{e}}^0= \mathrm{d}t-b_i \mathrm{d}x^i$, $\hat{\mathrm{e}}^\alpha=E^\alpha_{\hphantom{\alpha}i}\mathrm{d}x^i$
with $a_{ij}=\delta_{\alpha\beta}E^\alpha_{\hphantom{\alpha}i} E^\beta_{\hphantom{\beta}j}$. We  refer to this choice as the {\it Randers co-frame} and $b=b_i\mathrm{d}x^i$ as the {\it Randers one-form}.
The boundary data above describe a relativistic perfect fluid \footnote{Under certain kinematic conditions such as  shear- and expansion-less geodesic motion, the viscous component of the stress tensor vanishes, even for nonperfect fluids.} seen by a co-moving observer. The one-forms $\hat{\mathrm{e}}^a$ and their corresponding dual vector fields $\check{\mathrm{e}}_a$, $\hat{\mathrm{e}}^a(\check{\mathrm{e}}_b)={\delta^a}_b$, provide the observer's orthonormal frame with the metric given by the symmetric $(0,2)$-tensor $\hat g=\eta_{ab}\hat{\mathrm{e}}^a\otimes\hat{\mathrm{e}}^b$.  The stress tensor is a $(1,1)$-tensor $T=T^a_{\hphantom{a}b}\check{\mathrm{e}}_a\otimes \hat{\mathrm{e}}^b$, encapsulated in $\hat{f}^a$ as 
\beq
\label{fT}
\kappa \hat{f}^a=T(\hat{\mathrm{e}}^a,\cdot)=T^a_{\hphantom{a}b}\hat{\mathrm{e}}^b
\,,\quad \kappa=\frac{3M}{8\pi G_N L}\, .
\eeq
For a perfect fluid this tensor can be written as\be
\label{fluidemST}
T^a_{\hphantom{a}b} = (\varepsilon +p) u^au_b +p\delta^a_{\hphantom{a}b}
\ee
with $\varepsilon$ and $p$ the energy density and pressure. The normalized velocity field  $\check{u}=u^a\check{\mathrm{e}}_a$, $\eta_{ab}u^au^b=-1$, determines the fluid flow. 
Examining the Fefferman--Graham expansion for the above mentioned geometries we find
\beq
\label{efrel}
\hat{f}^0=-2\hat{\mathrm{e}}^0\,,\quad \hat{f}^\alpha=\hat{\mathrm{e}}^\alpha\,.
\eeq
Comparing (\ref{fT}), (\ref{fluidemST}) and (\ref{efrel}), we find $\varepsilon = 2p=2\kappa$: the Kerr and TN--AdS$_4$ geometries describe the  \emph{same conformal} fluid in different kinematical states. Furthermore, the velocity vector-field is $\check{u}=\check \mathrm{e}_0$, which shows that the frame $\check{\mathrm{e}}_a$ is co-moving.  This is a general result: holographic perfect fluids are such that their co-moving frame is the Randers frame. In this frame the properties of the rotating fluid, such as its vorticity, are  encoded entirely in the {\it leading} term in the Fefferman--Graham expansion.

The explicit expressions for Kerr are (see e.g. \cite{Pope})
\bea
\label{Kerr_Randers1}
&b = \frac{a}{\Xi}\sin^2\theta\ \mathrm{d}\phi\, , \quad
a_{ij}=L^2\mathrm{diag}\left(\frac{1}{\Delta_\theta},\frac{\Delta_\theta}{\Xi^2}\sin^2\theta\right)\, ,&\\
\label{Kerr_Randers3}
&\Delta_\theta=1-\alpha^2\cos^2\theta\,,\quad\Xi=1-\alpha^2\,,\quad \alpha=\nicefrac{a}{L}\, , &
\eea
where $a$ is the angular-velocity  parameter with $a<L$. For TN (see Eq. (\ref{TN-met}) with $z=\sin^2\nicefrac{\theta}{2}$)
\cite{Mann}) we have instead 
\beq
\label{TN_Randers}
b=-2n(1-\cos\theta) \mathrm{d}\phi\,,\quad a_{ij}=L^2\mathrm{diag}(1,\sin^2\theta)
\eeq
with $n$ the nut charge.
The boundary frame dual to the Randers co-frame is
\beq
\label{frameRanders}
\check{\mathrm{e}}_0 =  \partial_t\, ,\quad  \check{\mathrm{e}}_\alpha=E_\alpha^{\hphantom{\alpha}i}\left(b_i\partial_t+\partial_i\right)\,, \quad E_\alpha^{\hphantom{\alpha}i}E^\beta_{\hphantom{\beta}i}=\delta^\beta_\alpha\, .
\eeq
In the Randers geometry 
the integral lines of $\partial_t$
are geodesics: the fluid and the co-moving observers are inertial. Such observers can define the {\em fluid's physical surface} as the set of points which are synchronous events in the observer's frame whose tangent bundle is spanned by the vectors $\partial_i$ since $\mathrm{d}t( \partial_i)=0$. We can adopt as a basis an orthonormal combination  $\check{\mathrm{z}}_\alpha=L_\alpha^{\hphantom{\alpha}i}\partial_i$ with  $L_\alpha^{\hphantom{\alpha}i}L^\beta_{\hphantom{\beta}i}=\delta^\alpha_\beta$. 
Then, the parallel transport of the physical surface along $\check{\mathrm{e}}_0$ is the physical manifestation of the fluid's flow in the co-moving frame. We in fact find that the physical surface is not parallel transported along $\partial_t$, namely
\beq
\label{cov_alpha}
\nabla_{\partial_t}(\partial_i+b_i\partial_t)=\nabla_{\partial_t}\partial_i = \omega_{ij}a^{jk}\left(\partial_k+b_k\partial_t\right),
\eeq
with $\omega_{ij}$ the spacetime components of the vorticity form.
The latter is a two-form defined as 
$\omega =\frac{1}{2}\left(
\mathrm{d}\hat{u}
+\hat{u}\wedge \nabla_{\check{u}}\hat{u}
\right)$ 
and reduces here to $\frac{1}{2}\mathrm{d}b$ due to the absence of acceleration \footnote{The geodesic congruence tangent to $\partial_t$ has neither shear nor expansion, but only vorticity.}. Hence, the inertial observers perceive the fluid's flow as the rotation upon parallel transport of the geodesic congruence tangent to $\partial_t$. 
For the metrics above, its only non-zero components are along the spatial co-frame $\hat\mathrm{e}^\alpha$. We find for Kerr
\beq
\label{omega_Kerr}
\omega_{\mathrm{K}} = \frac{a}{L^2}\cos\theta \, \hat\mathrm{e}^1 \wedge \hat\mathrm{e}^2\,,
\eeq
which describes a cyclonic flow as seen from the co-moving frame. 
In the TN case we must be more careful. Noting that the globally defined one-form is $\hat{\mathrm{e}}^2$ rather than ${\mathrm{d}}\phi$, we see that the coefficient $b_2$ in (\ref{TN_Randers}) diverges at $\theta=\pi$. This induces a $\delta$-function singularity in the vorticity
\beq
\label{omega_TN}
 \quad\omega_{\mathrm{TN}}=-\frac{n}{L^2} \, \hat\mathrm{e}^1 \wedge \hat\mathrm{e}^2-\frac{n}{L^2}\delta_2(\theta-\pi)\,,
\eeq
where the last term denotes a singular two-form with support only at $\theta=\pi$. The normal part of  (\ref{omega_TN}) describes a vortex flow with constant vorticity.  
The $\delta$-function singularity is the boundary remnant of the Misner string \cite{misner:1963}, which extends, in the chosen coordinates radially along the $\theta=\pi$ axis, intersecting the boundary at a (neutral) ``Misner vortex''. Since we are not interested in compactifying the Lorentzian time coordinate, this string is physical, \cite{bonnor:1969, dowker:1974} and will have important consequences holographically. The $\delta$-function singularity noted above shows up either as a singular contribution to the torsion of a smooth connection, or equivalently, as a singular contribution to the Levi--Civita connection. 

Many interesting properties originate from the Misner string, among which the appearance of 
closed timelike curves (CTCs), extending out to the boundary of TN. The vectors tangent to the physical surface of the co-moving observer in the $2+1$ geometry,  $\check{X}=X^\a\check{\mathrm{z}}_\a =X^i\partial_i$ have norm $||\check{X}||^2 = X^iX^j(a_{ij}-b_ib_j)$. This is positive for Kerr ($a<L$), but can vanish or even become negative for TN: the norm of the vector $\partial_\phi$ for example is given by $1-\nicefrac{4n^2}{L^2}\, \tan^2\nicefrac{\theta}{2}$ and becomes null or timelike for $\theta\geq \theta_*$ with $\tan\nicefrac{\theta_*}{2}=\nicefrac{L}{2n}$. The corresponding tangent lines are thus CTCs, although not geodesics. The homogeneous nature of the TN boundary makes the full analysis of the properties regarding the CTCs quite subtle (see e.g. \cite{LPP, reboucas}). For the fluid at hand, in its specific kinematic state, these properties translate into the existence of a disk of angular opening $2(\pi-\theta_*)$ around the Misner vortex.  We will argue that inside this disk, the velocity of the fluid exceeds the local speed of light.
Hence, the TN fluid could be interpreted as a superluminally moving optical medium.  Since the bulk theory is such that the boundary does not have access to a charge current, the Misner vortex cannot be associated with a vortex in an ordinary superfluid, but is related to the spinning string of \cite{Mazur}.

The space spanned by the vectors $\check{\mathrm{z}}_\alpha$, {\it i.e.}
does not coincide with that spanned by the vectors $\check{\mathrm{e}}_\alpha$ orthogonal to $\check{\mathrm{e}}_0=\partial_t$.
It is then natural to ask what the normalized  timelike vector $\check{\mathrm{z}}_0$, orthogonal to $\check{\mathrm{z}}_\alpha$, is. Such a choice corresponds to $\check\mathrm{z}_a$ and $\check\mathrm{e}_a$ being related by a local Lorentz transformation. The congruences of $\check\mathrm{z}_0$ would be the worldlines of a different set of, generally non-inertial, observers \footnote{For this set of observers the space spanned by $\check\mathrm{e}_\alpha$ obeys $\mathrm{d}t(\check\mathrm{e}_\alpha)
-b(\check\mathrm{e}_\alpha)=0$. Since Fr\"obenius criterion is not fulfilled ($\mathrm{d} (\mathrm{d}t-b)=-2\omega \Leftrightarrow \left[b_i\partial_t+\partial_i, b_j\partial_t+\partial_j\right]= 2\omega_{ij}\partial_t$), it is not possible to define a universal time whose synchronous hypersurfaces, tangent to $\check\mathrm{e}_\alpha$, would be the fluid physical surfaces simultaneously for all these observers.}. 
We find for the frame and the dual co-frame: 
\bea
\label{frame_Zermelo}
&\check{\mathrm{z}}_0=\frac{1}{\gamma}\left(\partial_t+W^i\partial_i\right)\, , \quad  \check{\mathrm{z}}_\alpha=L_\alpha^{\hphantom{\alpha}i}\partial_i&\\
\label{coframe_Zermelo}
&\hat{\mathrm{z}}^0=\gamma \mathrm{d}t \, , \quad \hat{\mathrm{z}}^\alpha= L^\alpha_{\hphantom{\alpha}i}(\mathrm{d}x^i - W^i \mathrm{d}t)&
\eea
with
$\gamma^{-2}=1-a^{ij}b_i b_j\,,\quad W^i=-\gamma^2a^{ij}b_j.
$
In the new orthonormal frame, the boundary metric  reads 
\bea
\label{metric_Zermelo}
&\hat g= \frac{1}{\lambda}\left[-\mathrm{d}t^2 +h_{ij}\big(\mathrm{d}x^i -W^i \mathrm{d}t\big)\left(\mathrm{d}x^j -W^j \mathrm{d}t\right)\right]&\\
\label{h_lambda}
&h_{ij}=\lambda(a_{ij}-b_i b_j)=\lambda L^\alpha_{\hphantom{\alpha}i} L^\beta_{\hphantom{\beta}j}\delta_{\alpha\beta}  
\,,\quad \lambda \equiv\nicefrac{1}{\gamma^2}\, .&
\eea
This is the so-called Zermelo form of the metric (\ref{Randersmetric}) \cite{Gibbons}. 

The Zermelo frame  (\ref{frame_Zermelo}) is non-inertial with acceleration 
$\nabla_{\check{\mathrm{z}}_0}\check{u}=\frac{1}{\gamma}W^i\omega_{ij}a^{jk}L^\alpha_{\hphantom{\alpha}k}\check{\mathrm{z}}_\alpha$:
the Zermelo observers see a rotating fluid. For Kerr, the flow's velocity measured by a Zermelo observer has norm
 $||V||=\sqrt{\delta_{\a\b} V^\a V^\b}=\nicefrac{a\sin\theta}{L\sqrt{\Delta_\theta}}$, which is always bounded by 1. For TN we find $||V||=\nicefrac{2n}{L}\tan\nicefrac{\theta}{2}$, which exceeds unity exactly when $\theta>\theta_*$.  This coincides with the threshold mentioned previously and emerges alternatively as a singularity at $\theta=\theta_*$ in the local Lorentz transformation connecting the Randers to the Zermelo frame. Hence, the Kerr and TN fluids can be used to describe wave propagation in moving media \cite{LPP}. Since the fluids are conformal, sound waves will propagate with finite velocity and hence, both Kerr and TN fluids will exhibit acoustic horizons. On the other hand, electromagnetic wave propagation will encounter an optical horizon only in the TN fluid, exactly at the onset of the CTC region
 where the boundary fluid moves ``superluminally''. This is a physically sensible situation since the ``velocity of light'' in our boundary fluid, which is normalized to one, is generally smaller than the velocity of light in the vacuum. 
 The use of the Zermelo frame gives a physical \emph{raison d'\^etre} for the bulk CTCs of TN.

 For non-inertial frames it is appropriate to calculate Fermi derivatives along $\check{\mathrm{z}}_0$. We find \cite{LPP} that the Fermi acceleration vanishes for Kerr and the Zermelo frame coincides with the locally non-rotating ZAMO frame \cite{Bardeen}.
At each spacetime point where an inertial observer meets a non-inertial one, $W^i$ are the components of their relative velocity and $\gamma$ their relative Lorentz factor.  For Kerr we find 
$W^1=0 , W^2= -\nicefrac{a}{L^2}$,
which shows that the metric (\ref{metric_Zermelo}) can be made conformal to a static metric by a global Lorentz boost as was noticed in \cite{Minwalla}. This is no longer true for TN in which case the fluid's velocity reads (see (\ref{frame_Zermelo}))
$
\check{u}=\gamma\left(\check\mathrm{z}_0+V^\alpha\check\mathrm{z}_\alpha\right)$ with $ V^\alpha =-\frac{1}{\gamma}L^\alpha_{\hphantom{\alpha}i}W^i.
$

To close our discussion on the kinematical aspects of the holographic fluids, we note that general stationary metrics giving rise to a boundary metric of the Randers form (\ref{Randersmetric}) have a vorticity determined by $\mathrm{d}b$. The geodesic equation is equivalent to the equation for the orbits of a charged particle in a magnetic field, with the magnetic field given by the vorticity. This fact is most elegantly demonstrated in terms of the associated Finsler norm (see \cite{Gibbons} for details), but it is also already implicit in Eq.~(\ref{cov_alpha}). 

Besides giving rise to optical horizons, vortex flows are associated with Berry phase effects  \cite{Leonhardt}. We will here show that a holographic analysis of the TN geometry leads naturally to such effects. The simplest context for this is to consider a scalar field propagating on the bulk TN geometry, which is of the form  ($z=\sin^2\nicefrac{\theta}{2}$)
\bea
\mathrm{d}s^2&=&\frac{\mathrm{d}r^2}{V(r)}-V(r)\left(\mathrm{d}t+4nz\mathrm{d}\phi\right)^2\nonumber\\&+&(r^2+n^2)\left(\frac{\mathrm{d}z^2}{z(1-z)}+4z(1-z)\mathrm{d}\phi^2\right)\label{TN-met}
\eea
with
\bea
&V(r)=\frac{(r^2-n^2)(1+\frac{r^2}{L^2}+3\frac{n^2}{L^2})-2Mr+4n^2\frac{r^2}{L^2}}{r^2+n^2}\, .&
\eea
This has $SU(2)\times \mathbb{R}$ isometry generated by
\begin{equation}
\label{}
\begin{array}{rcl}
H&=&-i\pa_t,\quad L_3= -i(\pa_\phi-2n\pa_t),\\
L_\pm &=& \frac{ i\mathrm{e}^{\pm i\phi}}{\sqrt{z(1-z)}}\left(2nz\pa_t\mp i{z(1-z)}\pa_z+\frac{1-2z}{2}\pa_\phi\right)
\end{array}
\end{equation}
and clearly these extend to the boundary as well. The isometries act transitively for $\theta<\pi$ \footnote{The boundary of TN is a squashed three-sphere. The latter is a homogeneous space, but this property is invalidated globally (\emph{i.e.} at  $\theta = \pi$) for the universal covering required for non-compact time.}, and we should note that in general the orbits are not closed, but are of helical nature. 

Field fluctuations will naturally organize into representations of the isometry group; such representations will be labeled by the eigenvalues $\omega,m$ of $H$ and $L_3$ respectively, as well as by the $SU(2)$ quadratic Casimir. One finds that the Klein--Gordon equation of mass $\mu$ is fully separable and we may write a general solution as
\beqn
\Phi(r,t,z,\phi)=\sum_{m,\omega,\lambda} R_{\lambda,\omega}(r)Y_{\lambda,m,\Omega}(z) \mathrm{e}^{i(m-\Omega)\phi-i\omega t}\, ,
\eeqn
where $\Omega=2n\omega$ and the quadratic Casimir is given by $C=\lambda-\Omega^2$. Relevant to our discussion here is the angular equation (the radial equation is given in \cite{LPP}), which reads
\beqn
&&\pa_z\left[z(1-z)\pa_zY_{\lambda,m,\Omega}(z)\right]\nonumber \\
&&+\left[\lambda-\frac{(m+\Omega(2z-1))^2}{4z(1-z)}\right]Y_{\lambda,m,\Omega}(z)=0\,.
\eeqn
This is a hypergeometric equation whose general solutions are of the form
\beqn
\hspace{-.55cm}Y_{\lambda,m,\Omega}(z)&\sim& z^{\pm \frac{m-\Omega}{2}}(1-z)^{\pm \frac{m+\Omega}{2}} \nonumber \\
&&{}_2F_1(1+q\pm m,-q\pm m,1+m\mp\Omega;z)
\eeqn
(for brevity we have written $C=q(q+1)$). The geometry is smooth near $\theta=0$, so it is natural to require that the solutions be non-singular there. In fact, although for generic $m$ the solutions wind around the $\phi$ direction, if $Y_{\lambda,m,\Omega}(z\to 0)\to0$, then the solution will be single-valued around $\theta=0$. We would like to emphasize here  that this condition does not imply that the solutions are non-singular near $\theta=\pi$. In fact, non-singular solutions are obtained only if $\Omega,q,m$ are all (half-)integer. Requiring $\omega$ to be quantized is tantamount to requiring the time coordinate to be compact, or equivalently, that the Misner string be invisible. In non-compact time, as we stated above, the Misner string is physical, and in the holographic context we wish to be able to probe the system at arbitrary real frequency. 
We note that a general solution regular at $\theta=\pi$ will be non-single-valued around $\theta=\pi$, with monodromy given by $\exp{2\pi i(m-\Omega)}$. This can be interpreted as an anionic phase of the scalar in the presence of the Misner vortex \cite{LPP}.

To recap, the rotating fluids described here provide the relevant equilibrium states for the holographic description of two distinct physical systems: rotating atomic gases and analogue gravity systems. Rotation has already been used to simulate a magnetic field in a superconductor/superfluid using the charged Kerr--Newman--AdS$_4$ metric \cite{Sonner}. Nevertheless, many interesting physical systems are neutral and require the presence of uniform rotation \cite{cooper:2009} in which case the relevant bulk metric appears to be Lorentzian Taub--NUT AdS$_4$, which intriguingly also has a planar limit \cite{Mann}. 
For analogue gravity systems our holographic rotating fluids could be interpreted as the moving media through which sound or light waves propagate. 

To fully understand the holography of these models, it is necessary to consider their transport properties, which is a challenging problem that will be addressed elsewhere~\cite{LPP}. For Kerr--AdS, one expects to find quasi-normal modes corresponding to sound modes, and in general there will be a band around the equator in which the fluid velocity exceeds the sound velocity \cite{Barcelo} (but not the velocity of light) -- see also \cite{Shapere}. Thus one expects the presence of an analogue acoustic horizon. In the TN case, the situation is different: although one certainly expects an acoustic horizon, there is also an optical horizon  \cite{Cacciatori} beyond which the fluid velocity exceeds the speed of light (in the medium). This occurs, as we discussed above, beyond an angle $\theta_*$.  A detailed understanding of this requires the knowledge of transport properties in the boundary fluid, and in particular its refraction index.
It should also be noted that sonic and light propagation in moving media with vorticity is a highly non-trivial issue \cite{Stone}. Our work provides the means for a holographic extension of the results of the latter reference.

{\small
{\bf Acknowledgements:}

RGL is partially supported by the US Department of Energy under contract FG02-91-ER40709.
ACP is partially supported by the 
EU grant FP7-REGPOT-2008-1-CreteHEPCosmo-228644. PMP  acknowledges financial support by  the ERC 226371, the IFCPAR 4104-2, the ANR NT09-573739 and the ITN PITN-GA-2009-237920 grants.
We gratefully acknowledge the hospitality of the Galileo Galilei Institute in Florence, Italy. ACP and RGL thank the CPHT 
where parts of this work were done. }


\begin{thebibliography}{99}


\bibitem{Hartnoll:2008vx}
  S.A.~Hartnoll, C.P.~Herzog and G.T.~Horowitz,
  Phys.\ Rev.\ Lett.\  {\bf 101} (2008) 031601
  [arXiv:0803.3295 [hep-th]].

\bibitem{Faulkner:2010da}
  T.~Faulkner, N.~Iqbal, H.~Liu, J.~McGreevy and D.~Vegh,
  arXiv:1003.1728 [hep-th].

\bibitem{Dolan:2010mx}
  B.P.~Dolan,
  J.\ Phys. {\bf A44}, 175001 (2011)
  [arXiv:1011.6641 [cond-mat.str-el]].


\bibitem{Policastro:2001yc}
 G.~Policastro, D.T.~Son and A.O.~Starinets,
  Phys.\ Rev.\ Lett.\  {\bf 87}, 081601 (2001)
  [arXiv:hep-th/0104066].


\bibitem{Minwalla1}
  S.~Bhattacharyya, V.E.~Hubeny, S.~Minwalla and M.~Rangamani,
  JHEP {\bf 0802} (2008) 045
  [arXiv:0712.2456 [hep-th]].
  
  \bibitem{cooper:2009}
N.R. Cooper,  arXiv:0810.4398[cond-mat-.mes-hall];
N. Gemelke, E. Sarajlic, S. Chu,  arXiv:1007.2677[cond-mat.quant-gas]; 
M. Roncallia, M. Rizzi and J. Dalibard, arXiv:1105.5593[cond-mat.quant-gas]





\bibitem{Unruh}
  W.G.~Unruh,
  Phys.\ Rev.\ Lett.\  {\bf 46} (1981) 1351.

\bibitem{Leonhardt}
  U.~Leonhardt and P.~Piwnicki,
  Phys.\ Rev.\ Lett.\  {\bf 84} (2000) 822.

\bibitem{LPP}
R.G. Leigh, A.C. Petkou and P.M. Petropoulos, to appear.

\bibitem{Leigh:2007wf}
  R.G.~Leigh and A.C.~Petkou,
  JHEP {\bf 0711}, 079 (2007)
  [arXiv:0704.0531 [hep-th]].

\bibitem{Mansi:2008br}
  D.S.~Mansi, A.C.~Petkou and G.~Tagliabue,
  Class.\ Quant.\ Grav.\  {\bf 26}, 045008 (2009)
  [arXiv:0808.1212 [hep-th]].

\bibitem{Mansi:2008bs}
  D.S.~Mansi, A.C.~Petkou and G.~Tagliabue,
  Class.\ Quant.\ Grav.\  {\bf 26}, 045009 (2009)
  [arXiv:0808.1213 [hep-th]].

\bibitem{Carter}
  B.~Carter,
  Commun.\ Math.\ Phys.\  {\bf 10} (1968) 280.

\bibitem{Pope}
  G.W.~Gibbons, M.J.~Perry and C.N.~Pope,
  Class.\ Quant.\ Grav.\  {\bf 22} (2005) 1503
  [arXiv:hep-th/0408217].

\bibitem{Mann}
  D.~Astefanesei, R.B.~Mann and E.~Radu,
  JHEP {\bf 0501} (2005) 049
  [arXiv:hep-th/0407110].




\bibitem{Gibbons}
  G.W.~Gibbons, C.A.R.~Herdeiro, C.M.~Warnick and M.C.~Werner,
  Phys.\ Rev.\  {\bf D79} (2009) 044022
  [arXiv:0811.2877 [gr-qc]].




\bibitem{misner:1963}
C. Misner,
Jour. Math. Phys. \textbf{4} (1963) 924.

 \bibitem{bonnor:1969}
 W.B. Bonnor, 
 Proc. Camb. Phil. Soc. \textbf{66} (1975) 145.

\bibitem{dowker:1974}
J.S. Dowker, 
Gen. Rel. Grav.  \textbf{5} (1974) 603. 

\bibitem{reboucas}
F.M. Paiva, M.J. Reboucas and A.F.F. Texeira, Phys. Lett. \textbf{126A} (1987) 168.

\bibitem{Mazur}
  P.O.~Mazur,
  Phys.\ Rev.\ Lett.\  {\bf 57} (1986) 929.



\bibitem{Bardeen}
  J.M.~Bardeen, W.H.~Press and S.A.~Teukolsky,
  Astrophys.\ J.\  {\bf 178} (1972) 347.

\bibitem{Minwalla}
  S.~Bhattacharyya, R.~Loganayagam, S.~Minwalla, S.~Nampuri, S.P.~Trivedi and S.R.~Wadia,
  JHEP {\bf 0902}, 018 (2009)
  [arXiv:0806.0006 [hep-th]];
  M.M.~Caldarelli, O.J.C.~Dias and D.~Klemm,
  JHEP {\bf 0903} (2009) 025
  [arXiv:0812.0801 [hep-th]].







\bibitem{Sonner}
  J.~Sonner,
  Phys.\ Rev.\  {\bf D80}, 084031 (2009)
  [arXiv:0903.0627 [hep-th]].


\bibitem{Barcelo}
  C.~Barcelo, S.~Liberati and M.~Visser,
  Living Rev.\ Rel.\  {\bf 8} (2005) 12
  [arXiv:gr-qc/0505065].


\bibitem{Shapere}
  S.R.~Das, A.~Ghosh, J.H.~Oh and A.D.~Shapere,
  JHEP {\bf 1104} (2011) 030
  [arXiv:1011.3822 [hep-th]].


\bibitem{Cacciatori}
  F.~Belgiorno {\it et al.},
  arXiv:1009.4634 [gr-qc].


\bibitem{Stone}
  S.E.~Perez Bergliaffa, K.~Hibberd, M.~Stone and M.~Visser,
  arXiv:cond-mat/0106255.



\end{thebibliography}
\end{document}